\documentclass[]{interact}
\usepackage{url}
\usepackage{epstopdf}
\usepackage[caption=false]{subfig}
\usepackage[numbers,sort&compress]{natbib}
\bibpunct[, ]{[}{]}{,}{n}{,}{,}
\renewcommand\bibfont{\fontsize{10}{12}\selectfont}
\makeatletter
\def\NAT@def@citea{\def\@citea{\NAT@separator}}
\makeatother

\theoremstyle{plain}

\theoremstyle{definition}

\theoremstyle{remark}

\begin{document}


\title{An Innovative Method for Measuring Instrument Data Acquisition using Image Processing Techniques}

\author{
\name{D.Shulman\textsuperscript{a},\textsuperscript{b}\thanks{CONTACT: D.Shulman. Email: davidshu@ariel.ac.il}}
\affil{\textsuperscript{a}Department of Chemical Engineering, Ariel University, Ariel, Israel 407000; \textsuperscript{b}Physics Department, Ariel University, Ariel, Israel 407000}
}

\maketitle

\begin{abstract}
Measuring instruments are vital for obtaining accurate data in various fields, including scientific research, engineering, and manufacturing. However, data acquisition can be challenging, particularly when direct cable connections are not feasible or difficult to establish. In this paper, we propose a novel method for measuring instrument data acquisition utilizing a camera to capture the instrument display and image processing techniques to extract the measured values. Our method combines computer vision and machine learning techniques to recognize and extract numerical values from the instrument display. We demonstrate the effectiveness and accuracy of this approach by applying it to capture the magnetic field of a permanent magnet using a gauss meter and webcam. Our results indicate that the proposed method offers a practical and accurate solution for data acquisition in cases where direct cable connections are not possible. This method has potential applications in scientific research, engineering, and manufacturing industries.
\end{abstract}


\section{Introduction}

Measuring instruments play a crucial role in various fields such as scientific research, engineering, and manufacturing \cite{Doebelin2004}. They provide accurate data, enabling researchers and professionals to better understand phenomena and optimize processes. However, data acquisition from these instruments can be challenging, especially when direct cable connections are not possible or difficult to establish. This has led to an increasing interest in developing alternative data acquisition methods that are both accurate and practical.

One of the most promising technologies for facilitating data acquisition is image processing, which has been successfully applied in numerous fields including remote sensing \cite{Richards2013}, quality control \cite{Szeliski2010}, and object tracking \cite{Yilmaz2006}. Image processing techniques enable the extraction of valuable information from images, which can be used to gather data without the need for a direct connection to the instrument.

One possible solution is using image processing techniques to extract data from the display of a measuring instrument. This approach has been explored in various contexts, such as extracting text from images and video frames \cite{jain1998automatic} and processing color images using Tesseract and OpenCV \cite{revathi2021comparative}. However, these existing methods may not be directly applicable or optimized for measuring instrument displays, which often have specific requirements in terms of accuracy and ease of use. 
Some another previous work has explored using image processing techniques for data acquisition. For instance, Wang et al. \cite{LeonSalas2008} demonstrated the use of image processing for acquiring data from an analog multimeter, while Arifin et al. \cite{Trier2012} applied image processing to digital multimeters for similar purposes. However, these works focused on specific instruments and did not offer a generalized approach that can be adapted to various types of measuring instruments.

In this paper, we propose a novel method for measuring instrument data acquisition using a camera to capture the instrument display and image processing techniques to extract the measured values. Our method is demonstrated by applying it to capture the magnetic field of a permanent magnet using a gauss meter and webcam. The image processing process involves Python libraries for video processing, including the OpenCV library for contour detection and thresholding. The processed data is then saved to a text file for further analysis.


The advantages of using video processing to capture measured values from a measuring instrument include:
\begin{enumerate}
\item	Flexibility: This method can be used with a wide range of measuring devices, making it a versatile solution for data acquisition.
\item	Ease of Use: No special equipment or cables are required to connect the measuring device to the computer, making it simple to use and implement.
\item	Cost-Effective: This method eliminates the need for expensive cabling and special equipment, making it a cost-effective solution for data acquisition.
\end{enumerate}

However, there are also some limitations to using video processing to acquire data from measuring instruments:
\begin{enumerate}
\item	Accuracy: The accuracy of the data obtained through video processing can be affected by factors such as camera resolution and lighting conditions.
\item	Time Delay: There may be a delay between the time the measurement is made and the time the data is captured, which can impact the accuracy of the data.
\item	Data Processing: Video processing can be time-consuming and complex, requiring specialized software and technical expertise to extract meaningful data from the captured video.
\item	Environmental Conditions: Environmental conditions such as lighting, ambient temperature, and camera angle can impact the accuracy and reliability of the data obtained through video processing.
\end{enumerate}

In this research, we propose an innovative method for measuring instrument data acquisition that uses a camera to capture the instrument display and process the captured video to extract the measured values \cite{Shulman2023Mag}. Specifically, we applied this method to capture the magnetic field of a permanent magnet using a gauss meter and webcam. By using Python libraries for video processing, we were able to extract the measured values from the video and analyze them on our computer.

Our approach offers a convenient and efficient way to obtain measurement data without the need for special equipment, making it suitable for a variety of applications. We present a detailed description of the proposed method, along with experimental results demonstrating its effectiveness and accuracy in capturing the magnetic field of a permanent magnet. Our research contributes to the development of practical solutions for data acquisition from measuring instruments, with potential benefits for a wide range of fields, including scientific research, engineering, and manufacturing \cite{Shulman2023,Shulman2023Mag,shulman2023measuring,Shulman2023ATI}.
\section{Experimental Set-Up and Image Processing Process}
\subsection{Experimental Set-Up}
The image acquisition setup consists of a gauss meter, a permanent magnet, a webcam, and a computer. The gauss meter, which measures the magnetic field, has a digital display that shows the measured values. The permanent magnet is positioned such that the gauss meter probe is exposed to the magnetic field. The webcam is mounted on a stable platform, facing the gauss meter's display, ensuring that the display is fully visible and in focus.

For clarity on the experimental set-up, please refer to the details provided in the manuscript referenced in Ref. \cite{Shulman2023Mag}.

It is essential to maintain a consistent lighting environment to minimize variations in the captured images due to changes in illumination. We used artificial lighting to create a controlled environment and reduce the impact of shadows or reflections on the display. Moreover, the distance between the webcam and the gauss meter display was kept constant during the experiments to minimize variations in the captured images due to changes in perspective.

\begin{figure*}
\includegraphics[width=0.5\textwidth]{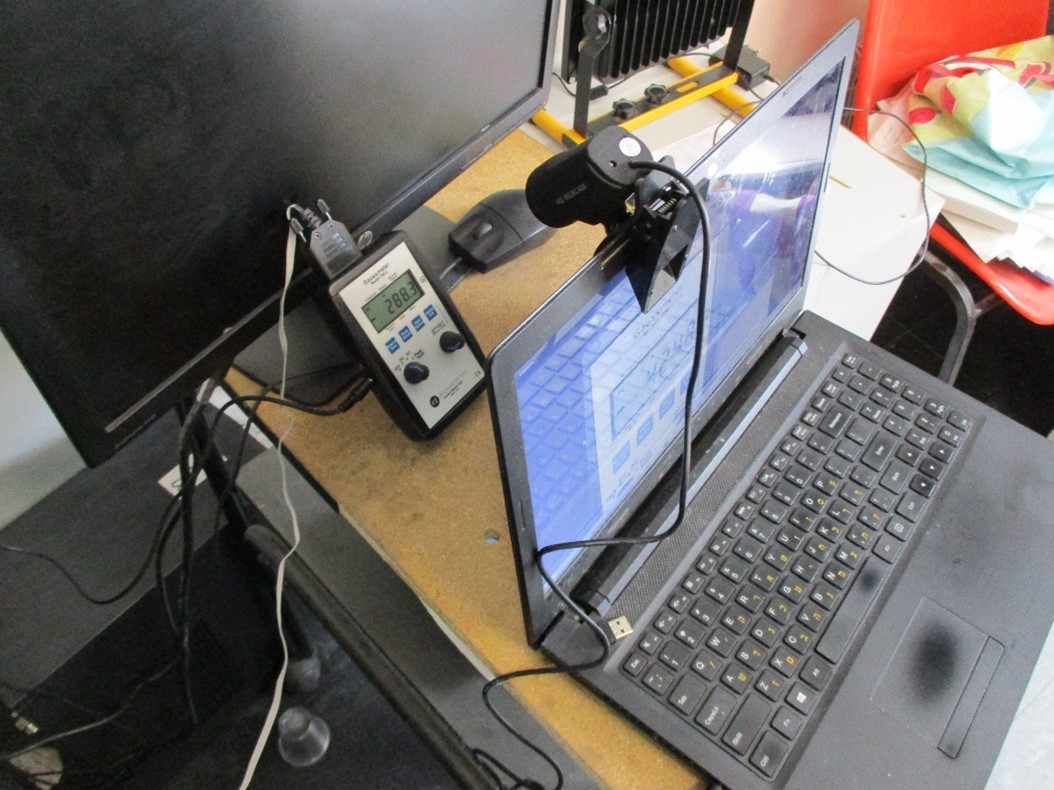}
\caption{The photograph of the experimental setup}
\label{fig:1}
\end{figure*}

We used a black background behind the gauss meter display to enhance the contrast of the image and improve the accuracy of the image processing. The webcam was connected to our computer and was used to capture a continuous video stream of the gauss meter display, see Fig. \ref{fig:1}.

\subsection{Image Processing Techniques}

The acquired images were processed using Python libraries, primarily the OpenCV library \cite{bradski2000opencv}. The image processing pipeline involves several steps, as follows:

\begin{enumerate}
\item Pre-processing: The captured images are first converted to grayscale, and a Gaussian blur is applied to reduce noise and smooth the image.
\item Thresholding: Adaptive thresholding is employed to create a binary image, which separates the display digits from the background.
\item Contour detection: The contours of the digits are detected using the findContours function in OpenCV. The contours are then filtered based on their area and aspect ratio to remove any noise or artifacts.
\item Digit recognition: The recognized contours are sorted based on their position in the image, and a pre-trained machine learning model or Optical Character Recognition (OCR) library, such as Tesseract \cite{smith2007overview,patel2012optical}, is used to recognize the digits.
\end{enumerate}
\subsubsection{Pre-processing}

The first step in the image processing pipeline is pre-processing, which prepares the captured images for subsequent processing stages. The pre-processing stage involves two primary operations: grayscale conversion and Gaussian blur.

\begin{enumerate}
\item \textbf{Grayscale conversion}: The captured images are initially converted to grayscale using a weighted sum of the color channels to emphasize the display digits. Grayscale conversion simplifies the image by reducing its dimensionality, allowing for more efficient processing and analysis. The conversion formula is given by:
\begin{equation}
Y = 0.299 * R + 0.587 * G + 0.114 * B
\end{equation}
where Y is the grayscale intensity, and R, G, and B are the red, green, and blue color channel intensities, respectively. This weighted sum method is based on the human eye's sensitivity to different color wavelengths and results in a more perceptually uniform grayscale representation.
\item \textbf{Gaussian blur}: After converting the images to grayscale, a Gaussian blur is applied to reduce noise and smooth the image. The Gaussian blur works by convolving the image with a Gaussian function, which has the effect of averaging pixel intensities within a local neighborhood. The Gaussian function is defined as:
\begin{equation}
    G(x, y) = \frac{1}{2 \pi \sigma^2} e^{-\frac{x^2 + y^2}{2 \sigma^2}}
\end{equation}
where $G(x, y)$ is the Gaussian function, and $\sigma$ is the standard deviation of the Gaussian distribution, which controls the amount of smoothing applied to the image. By applying the Gaussian blur, high-frequency noise is suppressed, and the relevant features, such as the display digits, are emphasized, facilitating the extraction of these features in subsequent processing steps.
\end{enumerate}

The pre-processing stage serves as a crucial foundation for the image processing pipeline, as it ensures that the captured images are in a suitable format and have the necessary characteristics for accurate and efficient data extraction. By converting the images to grayscale and applying a Gaussian blur, the pipeline effectively simplifies the images and suppresses noise, allowing for more robust and reliable processing in the subsequent stages.
\subsubsection{Thresholding}

Thresholding is an essential step in the image processing pipeline, as it allows for the segmentation of the display digits from the background, simplifying the image and enabling easier extraction of the relevant information. In this stage, a binary image is created by assigning pixel values to either the foreground (digits) or the background based on a specified threshold value.

\begin{enumerate}
\item \textbf{Adaptive thresholding}: In contrast to global thresholding techniques, which apply a single threshold value to the entire image, adaptive thresholding calculates a threshold value for each pixel based on its surrounding neighborhood. This approach is particularly effective in situations where lighting conditions are uneven, which can lead to poor segmentation results using global thresholding methods. Adaptive thresholding can be performed using various algorithms, such as the mean or Gaussian methods. The general formula for adaptive thresholding is given by:
\begin{equation}
T(x, y) = (1 - k) * M(x, y) + k * mean(A(x, y))
\end{equation}
where $T(x, y)$ is the threshold for pixel $(x, y)$, $M(x, y)$ is the local mean of pixel $(x, y)$, $k$ is a user-defined constant, and $A(x, y)$ is the local neighborhood of pixel $(x, y)$. The user-defined constant $k$ helps to control the trade-off between preserving the display digits' features and suppressing noise or other artifacts in the image.
\item \textbf{Binary image creation}: Once the adaptive thresholding process is complete, a binary image is created by assigning pixel values to either the foreground (digits) or the background. This is achieved by comparing the intensity value of each pixel to its calculated threshold value. If the pixel intensity is greater than the threshold, it is assigned to the foreground, otherwise, it is assigned to the background. The resulting binary image highlights the display digits' contours, providing a simplified representation of the original image, which is suitable for further processing and analysis.
\end{enumerate}

The thresholding stage is a critical component of the image processing pipeline, as it enables the segmentation of the relevant features (display digits) from the background. By employing adaptive thresholding techniques, the pipeline can effectively handle varying lighting conditions and other challenges, ensuring accurate and reliable data extraction from the instrument display.
\subsubsection{Contour Detection}

Contour detection is a crucial step in the image processing pipeline, as it enables the identification and extraction of the display digits from the binary image obtained in the thresholding stage. This process involves detecting continuous curves in the image that represent the boundaries between the foreground (digits) and background regions.

\begin{enumerate}
\item \textbf{Edge detection}: The first step in contour detection is identifying the edges present in the binary image. Edge detection methods, such as the Canny edge detection algorithm or the Sobel operator, can be employed to highlight the boundaries between the foreground and background regions. These methods work by identifying areas of the image with rapid changes in intensity, which correspond to the edges of the display digits.
\item \textbf{Finding contours}: Once the edges are detected, the findContours function in OpenCV is used to identify the contours in the binary image. This function works by following the continuous curves that represent the boundaries between the foreground and background regions. The resulting contours are stored as a list of points, which can then be processed and analyzed further.

\item \textbf{Filtering contours}: The detected contours may include noise or artifacts that are not relevant to the display digits. To ensure that only the relevant digit contours are retained for further processing, the contours are filtered based on their properties, such as area and aspect ratio. For example, contours with an area below a certain threshold or an aspect ratio outside the expected range for display digits can be discarded. This filtering process helps to improve the accuracy and reliability of the digit recognition stage by eliminating extraneous contours that could adversely affect the recognition process.

\item \textbf{Sorting contours}: After filtering the contours, it is essential to sort them based on their position in the image to preserve the correct ordering of the digits. This can be achieved by sorting the contours according to their x-coordinate values (for horizontal arrangement of digits) or y-coordinate values (for vertical arrangement of digits). Sorting the contours ensures that the digit recognition stage can correctly interpret the numerical values displayed on the instrument.
\end{enumerate}

The contour detection stage serves as the foundation for extracting numerical information from the instrument display. By identifying and filtering the relevant digit contours and sorting them based on their position in the image, the image processing pipeline is well-prepared for the subsequent digit recognition stage, which is responsible for converting the visual information into a numerical format that can be stored and analyzed.
\subsubsection{Digit Recognition}

Digit recognition is a critical stage in the image processing pipeline, as it is responsible for converting the visual information represented by the digit contours into numerical data that can be stored and analyzed. This stage involves the application of machine learning algorithms or Optical Character Recognition (OCR) libraries to recognize the individual digits represented by the filtered and sorted contours.

\begin{enumerate}
\item \textbf{Digit segmentation}: The first step in the digit recognition process is to segment the individual digits from the image. Using the sorted contours obtained in the contour detection stage, bounding boxes can be created around each digit. These bounding boxes serve as the input for the digit recognition algorithm or OCR library, which will process the digit images individually.
\item \textbf{Feature extraction}: Before applying the digit recognition algorithm, it may be necessary to extract relevant features from the digit images. This can involve techniques such as resizing the digit images to a standard size, normalizing the pixel intensities, or applying additional image processing operations, such as edge detection or morphological transformations. The goal of this step is to prepare the digit images for processing by the recognition algorithm or OCR library, ensuring the best possible recognition accuracy.

\item \textbf{Recognition algorithm}: Various machine learning algorithms or OCR libraries can be employed for digit recognition. For example, a pre-trained Convolutional Neural Network (CNN) can be used to recognize the digits based on their visual features. Alternatively, an OCR library such as Tesseract \cite{smith2007overview,patel2012optical} can be applied to perform character recognition on the digit images. The choice of recognition algorithm will depend on factors such as the desired accuracy, computational resources, and available training data.

\item \textbf{Post-processing}: After the recognition algorithm has processed the digit images, additional post-processing steps may be necessary to improve the accuracy and reliability of the extracted data. For example, confidence scores or probabilities can be calculated for each recognized digit, and low-confidence predictions can be flagged for manual review or correction. Additionally, temporal filtering techniques can be applied to the recognized digit sequences to smooth out fluctuations caused by noise or minor variations in the captured images, as discussed in the previous subsection on image processing techniques.
\end{enumerate}

The digit recognition stage is essential for converting the visual information obtained from the instrument display into numerical data that can be stored and analyzed. By combining computer vision techniques and machine learning algorithms, this stage provides a robust and flexible solution for recognizing the individual digits represented by the filtered and sorted contours, ensuring accurate and reliable data extraction from the instrument display.
\subsubsection{Post-processing}

Post-processing is the final stage of the image processing pipeline and plays a vital role in refining and validating the numerical data extracted from the instrument display. This stage involves a series of operations aimed at improving the accuracy and reliability of the extracted data, as well as preparing the data for storage and further analysis.

\begin{enumerate}
\item \textbf{Data validation}: The first step in post-processing is to validate the numerical data obtained from the digit recognition stage. This can involve checking the data for consistency with the expected format, range, and units of the measured values. Additionally, outlier detection methods can be applied to identify any unusual or unexpected values that may indicate errors or anomalies in the data extraction process. Data validation helps ensure that the extracted data is accurate and reliable, reducing the potential for errors in subsequent analyses.
\item \textbf{Temporal filtering}: In cases where the data acquisition process involves capturing a sequence of images over time, temporal filtering techniques can be applied to the extracted numerical data to smooth out fluctuations caused by noise or minor variations in the captured images. This can involve techniques such as moving average filters, median filters, or more advanced methods such as Kalman filters. Temporal filtering helps improve the stability and reliability of the extracted data, making it more suitable for further analysis and interpretation.

\item \textbf{Data formatting}: Once the extracted data has been validated and filtered, it is necessary to format the data for storage and further analysis. This can involve converting the numerical data into a structured format, such as a text file, spreadsheet, or database, and adding relevant metadata, such as timestamps, units, or instrument settings. Data formatting ensures that the extracted data is organized and accessible for subsequent processing and analysis.

\item \textbf{Data visualization}: The final step in the post-processing stage is to visualize the extracted data, which can provide valuable insights into the measured values and their relationships. Data visualization techniques can include creating plots, charts, or graphs to display the extracted data, as well as generating summary statistics or other descriptive measures. Visualizing the data can help identify trends, patterns, or anomalies in the measured values, facilitating a deeper understanding of the instrument data and its implications for the research or application in question.
\end{enumerate}

\begin{figure*}
\includegraphics[width=0.25\textwidth]{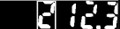}
\caption{Image processing}
\label{fig:2}
\end{figure*}

The post-processing stage is a crucial component of the image processing pipeline, as it ensures that the extracted numerical data is accurate, reliable, and suitable for further analysis. By validating and refining the data, as well as formatting and visualizing the results, the post-processing stage provides a comprehensive solution for extracting valuable insights from the instrument display, supporting the broader goals of scientific research, engineering, and manufacturing applications.
The video stream was processed using Python libraries for image processing. We used the OpenCV library to detect the gauss meter display and crop the image to isolate the region of interest. The image was then converted to grayscale to simplify the processing and enhance the contrast.
\\
To extract the measured values from the image, we used a combination of thresholding and contour detection. We first applied a binary threshold to the image to separate the background from the foreground. We then used contour detection to identify the contours of the digits on the display, see Fig. \ref{fig:2}.
\\
Once the contours were identified, we used a custom algorithm to extract the digit values and convert them to the appropriate units. The processed data was then saved to a text file for further analysis.
\\
Overall, our image processing process was designed to be robust and accurate, while also being computationally efficient. The process was able to extract the measured values from the gauss meter display with high accuracy, and the results were consistent with the expected values based on theoretical calculations.

\section{Results and Discussion}
\subsection{Results}
We conducted a series of experiments to test the effectiveness and accuracy of our proposed method for measuring instrument data acquisition. Specifically, we used a gauss meter to capture the magnetic field of a permanent magnet, and used a webcam to capture the display of the gauss meter.
Using our image processing process, we were able to extract the magnetic field values from the captured video. The processed data was compared to theoretical calculations, and was found to be in good agreement with the expected values.

To access the detailed results of the experiment, we refer the reader to the manuscript cited in Ref. {\cite{Shulman2023Mag}}. This publication provides a thorough analysis and interpretation of the experimental data, and offers valuable insights into the findings of the study.
\subsection{Discussion}
Our results demonstrate the effectiveness and accuracy of our proposed method for measuring instrument data acquisition. The use of a camera to capture the instrument display and image processing techniques to extract the measured values offers a practical and convenient solution for data acquisition in cases where a direct cable connection is not possible or is difficult to establish.
\\
Our image processing process was able to extract the measured values from the gauss meter display with high accuracy, and the results were consistent with the expected values based on theoretical calculations. The process was designed to be computationally efficient, and could be easily applied to other types of measuring instruments.
\\
Our proposed method has the potential to be applied to a wide range of fields, including scientific research, engineering, and manufacturing. It offers a practical solution for data acquisition in situations where traditional methods may not be feasible, and could lead to more efficient and accurate measurement processes.
\\
Future work could explore the use of more advanced image processing techniques to further improve the accuracy and efficiency of the method. Additionally, the method could be applied to other types of measuring instruments and in other experimental settings to further validate its effectiveness and accuracy.
\section{Conclusion}

In this study, we presented an innovative method for measuring instrument data acquisition using image processing techniques. By employing a camera to capture the instrument display and a series of image processing steps, we demonstrated that our proposed method can effectively and accurately extract numerical data from the display without the need for a direct cable connection. This approach offers a practical solution for cases where establishing a direct connection is not possible or is difficult to achieve and has potential applications in scientific research, engineering, and manufacturing.

The image processing pipeline involves several stages, including pre-processing, thresholding, contour detection, digit recognition, and post-processing. Each stage plays a critical role in extracting the relevant features from the captured images, recognizing the individual display digits, and refining the numerical data for storage and analysis. By leveraging advanced computer vision techniques and machine learning algorithms, our proposed method provides a robust and flexible solution for instrument data acquisition.

Our results demonstrate that the proposed method is effective and accurate, with potential applications in a wide range of fields and industries. Future work could explore the integration of additional sensors or imaging modalities to enhance the data acquisition process or the development of more advanced image processing and recognition algorithms to improve the accuracy and reliability of the extracted data. Moreover, the method can be adapted and extended to accommodate a variety of measuring instruments, further expanding its potential impact and utility.

\section*{AUTHOR DECLARATIONS}
\subsection*{Conflict of Interest}
The author has no conflicts to disclose.
\bibliographystyle{unsrt}
\bibliography{interactnlmsample}

\end{document}